\documentclass[structabstract]{aa}
\usepackage{graphicx}
\usepackage{txfonts}
\begin{document}
   \title{Hip~63510C, Hip~73786B, and nine new isolated high proper motion 
          T dwarf candidates from UKIDSS DR6 and SDSS DR7} 


\titlerunning{Hip~63510C, Hip~73786B, and nine new isolated high proper motion
          T dwarf candidates}

   \author{R.-D. Scholz
          }
   \institute{Astrophysikalisches Institut Potsdam,
              An der Sternwarte 16, 14482 Potsdam, Germany\\
              \email{rdscholz@aip.de}
             }

   \date{Received 16 February, 2010; accepted 18 March, 2010}


  \abstract
   {}
   {Completing the poorly known substellar census of the solar neighbourhood,
    especially with respect to the coolest 
    brown dwarfs, will lead to a better
    understanding of failed star formation processes and binary
    statistics with different environmental conditions.
   }
   {Using UKIDSS data and their cross-correlation with the SDSS,
    we searched for high proper motion mid- to late-T dwarf candidates 
    with extremely blue near-infrared ($J$$-$$K$$<$0) and very red 
    optical-to-near-infrared ($z$$-$$J$$>$$+$2.5) colours.
   }
   {With 11 newly found T dwarf candidates, the proper motions of
    which range between 100 and 800 mas/yr, we increased the number 
    of UKIDSS T dwarf discoveries by $\approx$30\%. 
    Large proper motions were also measured
    for six of eight previously known T4.5-T9 dwarfs detected in our survey.
    All new candidates can be classified as T5-T9 dwarfs based on their 
    colours. Two of these objects were found to be common
    proper motion companions of Hipparcos stars with accurate parallaxes.
    The latter allow us to determine absolute magnitudes from which we
    classify Hip~63510C as T7 and Hip~73786B as T6.5 
    dwarfs
    with an uncertainty of $\pm$1 spectral subtype.
    The projected physical separation from their low-mass (M0.5 and K5) 
    primaries is in both cases
    about 1200~AU. One of the Hipparcos stars has already a known 
    very low-mass star or brown dwarf companion
    on a close astrometric orbit (Hip~63510B = Gl~494B).
    With distances of only 11.7 and 18.6~pc, deduced from their primaries
    respectively for 
    Hip~63510C and Hip~73786B, various follow-up observations can easily
    be carried out to study these cool brown dwarfs in more detail and
    to compare their properties with those of the already well-investigated 
    primaries.
   }
   {}


   \maketitle
%

\section{Introduction}

New deep near-infrared surveys like
UKIDSS\footnote{The UKIDSS project is defined in 
Lawrence et al.~(\cite{lawrence07}).
UKIDSS uses the UKIRT Wide Field Camera (WFCAM; Casali et al.~\cite{casali07})
and a photometric system described in Hewett et al.~(\cite{hewett06})
which is situated in the Mauna Kea Observatories (MKO) system (Tokunaga et
al.~\cite{tokunaga02}) The pipeline processing and science archive are described
in Irwin et al.~(\cite{irwin10}) and Hambly et al.~(\cite{hambly08}).} and
the Canada-France Brown Dwarf Survey (CFBDS; Delorme et al.~\cite{delorme08b}) 
provide a powerful tool for 
detecting even more of the coolest brown dwarfs (mid- and late-T dwarfs) in the
solar neigbourhood than previously found in the 
SDSS (Abazajian et al.~\cite{abazajian09})
and 2MASS (Skrutskie et al.~\cite{skrutskie06}).
From 155 currently known T dwarfs (see Gelino et al.~\cite{gelino09}
and references therein), there were 52 discovered with SDSS and 49 with 2MASS
during the last 12 years. The number of
discoveries from the UKIDSS large area survey (LAS) 
has already grown to 33 in only three years,
where the majority were found by Lodieu et al.~(\cite{lodieu07})
and Pinfield et al.~(\cite{pinfield08}).
The latest-type objects discovered in the SDSS and
2MASS are of spectral type T7 (Chiu et al.~\cite{chiu06}) and 
T8 (Burgasser et al.~\cite{burgasser02};
Tinney et al.~\cite{tinney05}; Looper, Kirkpatrick, \&
Burgasser~\cite{looper07}) respectively, whereas a few 
T8.5-T9 dwarfs have been found in UKIDSS
(Warren et al.~\cite{warren07};
Burningham et al.~\cite{burningham08,burningham09}) and
CFBDS (Delorme et al.~\cite{delorme08a}). 

Since most of the stellar neighbours of the Sun were originally
detected as high proper motion stars, one can expect to find
the nearest cool brown dwarfs in new high proper motion surveys
using deep multi-epoch near-infrared imaging
data over large areas of sky, as e.g. eventually provided
by two epochs of $J$-band data in UKIDSS. However, one can
already try to use the SDSS (Abazajian et al.~\cite{abazajian09}),
the deepest available optical survey overlapping
with the ongoing UKIDSS LAS, as the first epoch. 
The nearest cool brown dwarfs may still be visible in the red optical $z$-band 
of the SDSS data observed a few years before the UKIDSS
observations started.

We describe a high proper motion survey for nearby T dwarfs using
mainly UKIDSS and SDSS data which led to the discovery of
11 new mid- to late-T dwarf candidates, including two 
common proper motion objects of nearby red dwarf stars with
accurate Hipparcos parallaxes. 

\section{Selection of T dwarf candidates from UKIDSS} 

For the selection of T dwarf candidates, we used the available
cross-matching with the SDSS DR7 (Abazajian et al.~\cite{abazajian09})
as provided in the UKIDSS data base. To select faint objects 
that appear very red in optical-to-near-infrared colours, 
but blue in the near-infrared, 
as expected for mid- to late-T dwarfs, we used the following criteria:\\

$J1 > 11$,~~~~~~~ $J1-K < 0$,~~~~~~~ $z-J1 > +2.5$,\\

where $J1$ are the first epoch $J$-band data, and $z$ comes from
the nearest matching SDSS object within the search radius of 
10~arcsec. By doing so, we allowed for missing $Y$- and $H$-band 
data. 
Whereas our main search criterion was the negative $J$$-$$K$ colour,
we excluded all kind of optically blue objects (early-type stars and
white dwarfs) with the $z$$-$$J$ colour cut. Late-M and L dwarfs
would also meet the $z$$-$$J$ colour but not the $J$$-$$K$ colour criterion. 
L subdwarfs (e.g. Lodieu et al.~\cite{lodieu10}) have
negative $J$$-$$K$ but bluer $z$$-$$J$ ($\approx$$+$1.5). However, low-metallicity
T dwarfs would possibly also satisfy our colour cuts, as the peculiar T6
dwarf 2MASS~J0937$+$29 (Burgasser et al.~\cite{burgasser06a}) has 
$z$$-$$J$$\approx$+3.5 (derived from SDSS DR7 and Leggett et al.~\cite{leggett10}
photometry).
Note that a large $i$$-$$z$ was not required as a pre-condition.
4940 targets were found satisfying these criteria.

%
\begin{table*}[t]
\caption{SDSS DR7 $z$ and UKIDSS DR6 $YJHK$
photometry for new T dwarf candidates}
\label{tab_photom}
\centering
\begin{tabular}{l l l l l l l l }
\hline\hline
object  &   $z$       &  $Y$    & $J$     & $H$     & $K$     \\
\hline
ULAS~J032920.22$+$043024.5  & 20.14$\pm$0.13$^*$$^{\dagger}$   & 18.67$\pm$0.05     & 17.50$\pm$0.03     & 17.88$\pm$0.07     & 18.17$\pm$0.19     \\
ULAS~J081918.58$+$210310.8  & 20.73$\pm$0.21         & 18.25$\pm$0.03     & 16.95$\pm$0.01$^*$ & 17.28$\pm$0.04     & 17.18$\pm$0.06     \\
ULAS~J094516.39$+$075545.6  & 20.34$\pm$0.20         & 18.71$\pm$0.03$^*$ & 17.54$\pm$0.02$^*$ & 17.76$\pm$0.04$^*$ & 17.84$\pm$0.07$^*$ \\
ULAS~J101243.53$+$102101.6  & 20.22$\pm$0.20         & 18.02$\pm$0.03     & 16.88$\pm$0.01     & 17.25$\pm$0.05     & 17.45$\pm$0.08     \\
ULAS~J130041.72$+$122114.7  & 20.22$\pm$0.19         & 17.72$\pm$0.03     & 16.69$\pm$0.02     & 17.01$\pm$0.04     & 16.90$\pm$0.06     \\
ULAS~J141756.22$+$133045.8  & 20.42$\pm$0.17         & 17.94$\pm$0.03     & 16.77$\pm$0.01     & 17.00$\pm$0.03     & 17.00$\pm$0.04     \\
ULAS~J144901.90$+$114711.3  & 20.16$\pm$0.16         & 18.35$\pm$0.04     & 17.36$\pm$0.02     & 17.73$\pm$0.07     & 18.10$\pm$0.15     \\
ULAS~J150457.65$+$053800.8  & 19.88$\pm$0.13         & 17.65$\pm$0.02     & 16.59$\pm$0.02     & 17.05$\pm$0.04     & 17.41$\pm$0.09     \\
ULAS~J232035.28$+$144829.8  & 20.08$\pm$0.18         & 17.94$\pm$0.03     & 16.76$\pm$0.02     & 17.11$\pm$0.04     & 17.25$\pm$0.10     \\
ULAS~J232123.79$+$135454.8  & 20.11$\pm$0.17         & 17.65$\pm$0.03     & 16.69$\pm$0.02     & 17.09$\pm$0.06     & 17.36$\pm$0.10     \\
ULAS~J234228.96$+$085620.1  & 20.00$\pm$0.13         & 17.42$\pm$0.02     & 16.37$\pm$0.01     & 16.73$\pm$0.03     & 16.98$\pm$0.07     \\
\hline
\end{tabular}

\smallskip

\scriptsize{
Notes: $^*$ = mean values obtained from multiple measurements.
$^{\dagger}$ From two measurements $z$$=$20.75$\pm$0.20 and $z$$=$19.53$\pm$0.18,
the latter is problematic (different from PSF magnitude).
$z$ magnitudes are on the AB system.
$YJHK$ magnitudes are {\it{aperMag3}} derived for point sources
(Dye et al.~\cite{dye06}) and are on the
Vega system using the MKO photometric system.
}
\end{table*}

UKIDSS $Y, J1, J2, H$, and $K$ finding charts of all these candidates
were inspected to exclude the vast majority of them consisting
of ghost images of bright stars, images affected by
diffraction spikes, satellite trails, galaxies, and asteroids.
For the remaining star-like sources, the  
corresponding SDSS finding charts and DR7 data were checked
for possible counterparts
We have also used the SDSS database available at
Princeton (Finkbeiner et al.~\cite{finkbeiner04}),
where we found some additional epochs
for determining the proper motions.
Only 11 new candidates with a unique very red
SDSS counterpart 
(detected only in the $z$-band with $z$$\approx$20, whereas $i$$>$23
values listed in the SDSS DR7 indicate that there was no flux above the noise)
within a smaller search radius of 5~arcsec and a
significant proper motion were finally selected and are presented in
Sect.~\ref{sect_pmphot}. 
Note that these counterparts cannot be background late-M or L dwarfs,
since they would show up in UKIDSS with $J$$\approx$18 and $K$$\approx$16-17,
which is not the case.
In addition, eight previously known T4.5-T9 dwarfs were detected
and proper motions determined for six of them with identified
SDSS counterparts (Sect.~\ref{sect_knownT}).

We have also checked 2MASS, DENIS (Epchtein et al.~\cite{epchtein97})
and SSS (Hambly et al.~\cite{hambly01}) $I$-band data
for possible counterparts but failed to find any for the
new T dwarf candidates due to their faintness.

   \begin{figure}[b]
   \centering
   \includegraphics[height=39mm,angle=0]{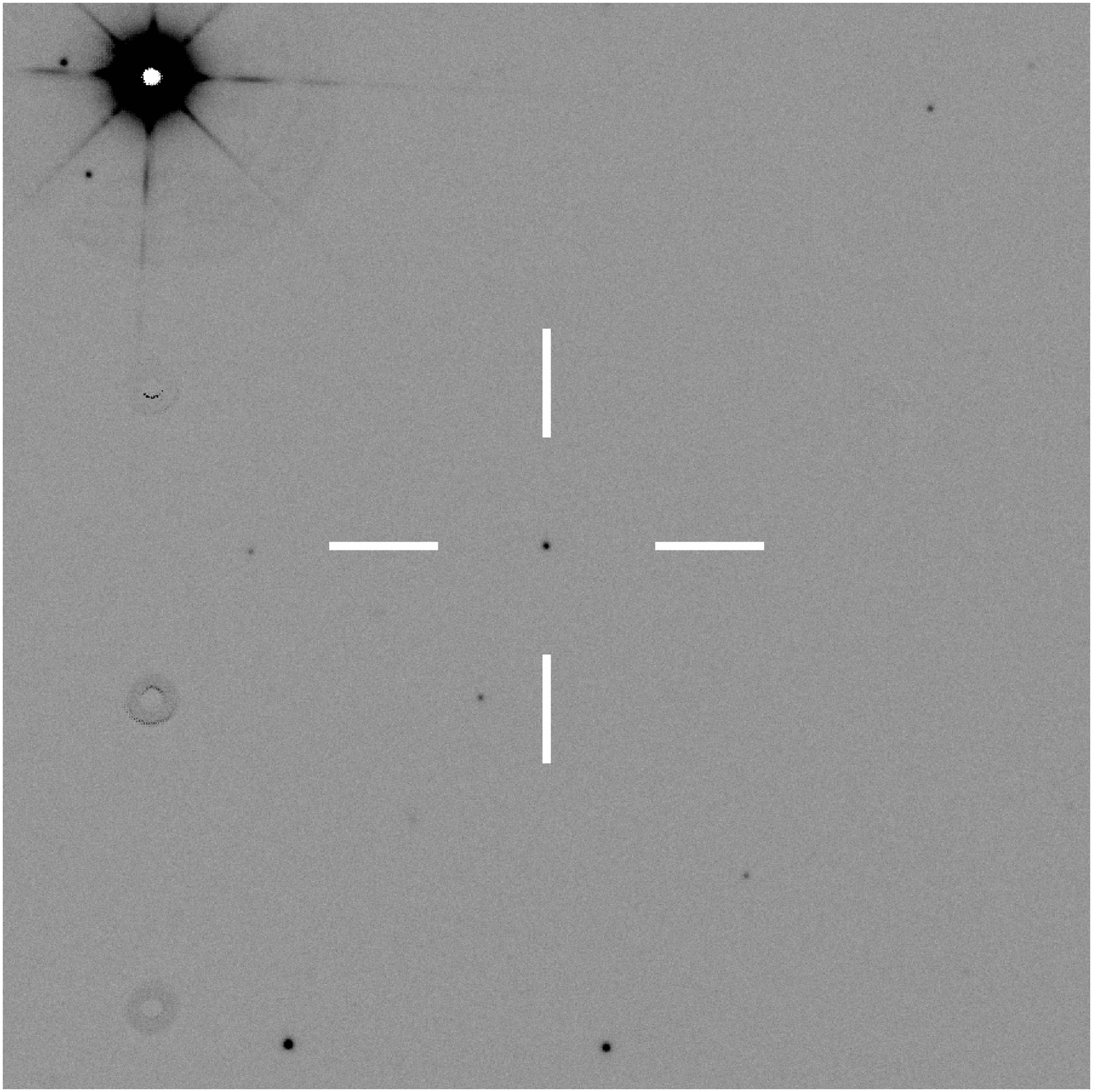}
   \includegraphics[height=39mm,angle=0]{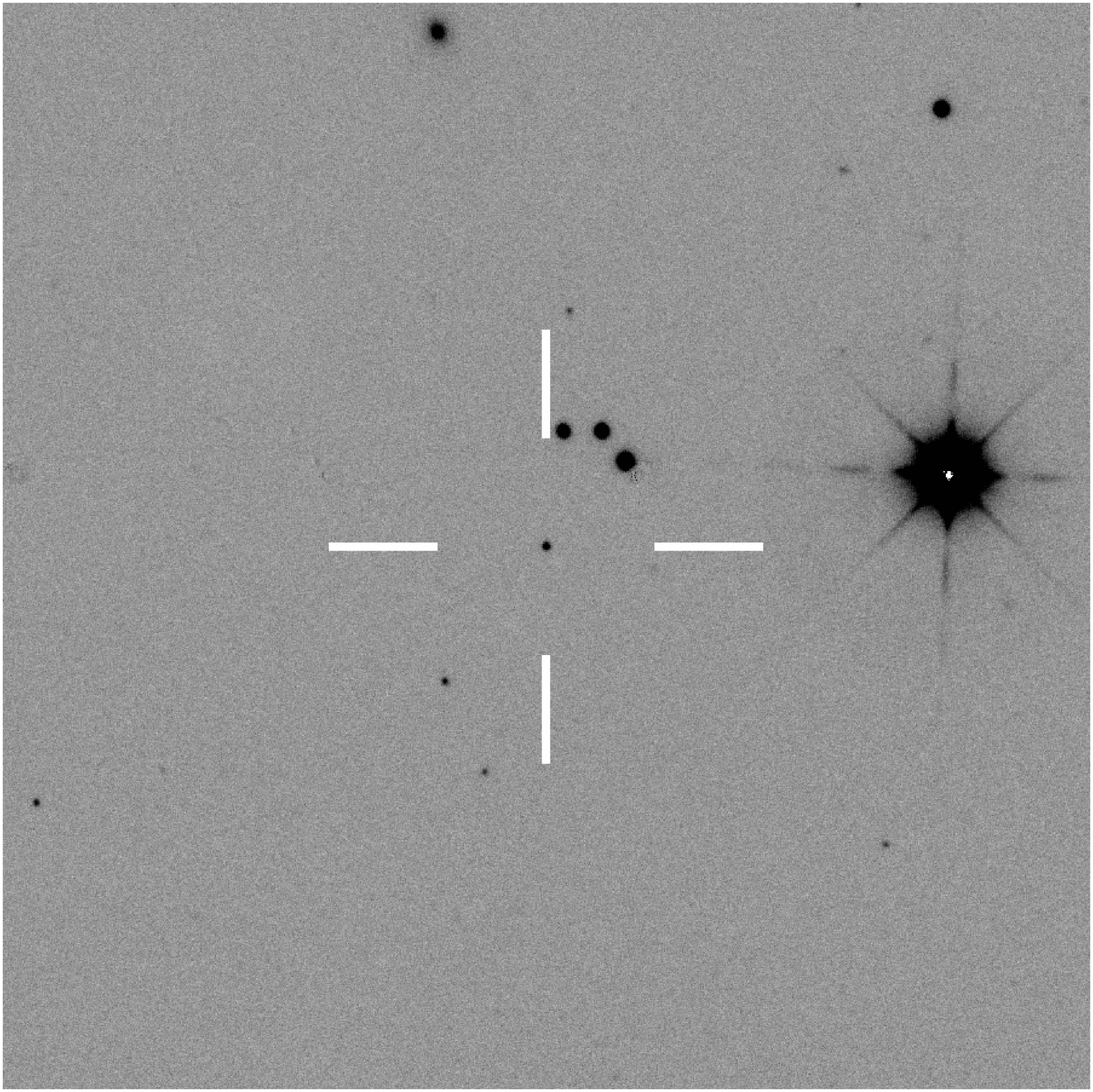}
      \caption{3$\times$3 arcmin$^2$ UKIDSS $J$-band images, left:
              ULAS~J1300$+$12 (= Hip~63510C) with the bright primary,
              Hip~~63510A in the upper left corner, right:
              ULAS~J1504$+$05 (= Hip~73786B) with its primary,
              Hip~73786A on the right side.
              North is up, East to the left.
              }
         \label{fig_fcharts2}
   \end{figure}

   \begin{figure}[t]
   \centering
   \includegraphics[height=26mm,angle=0]{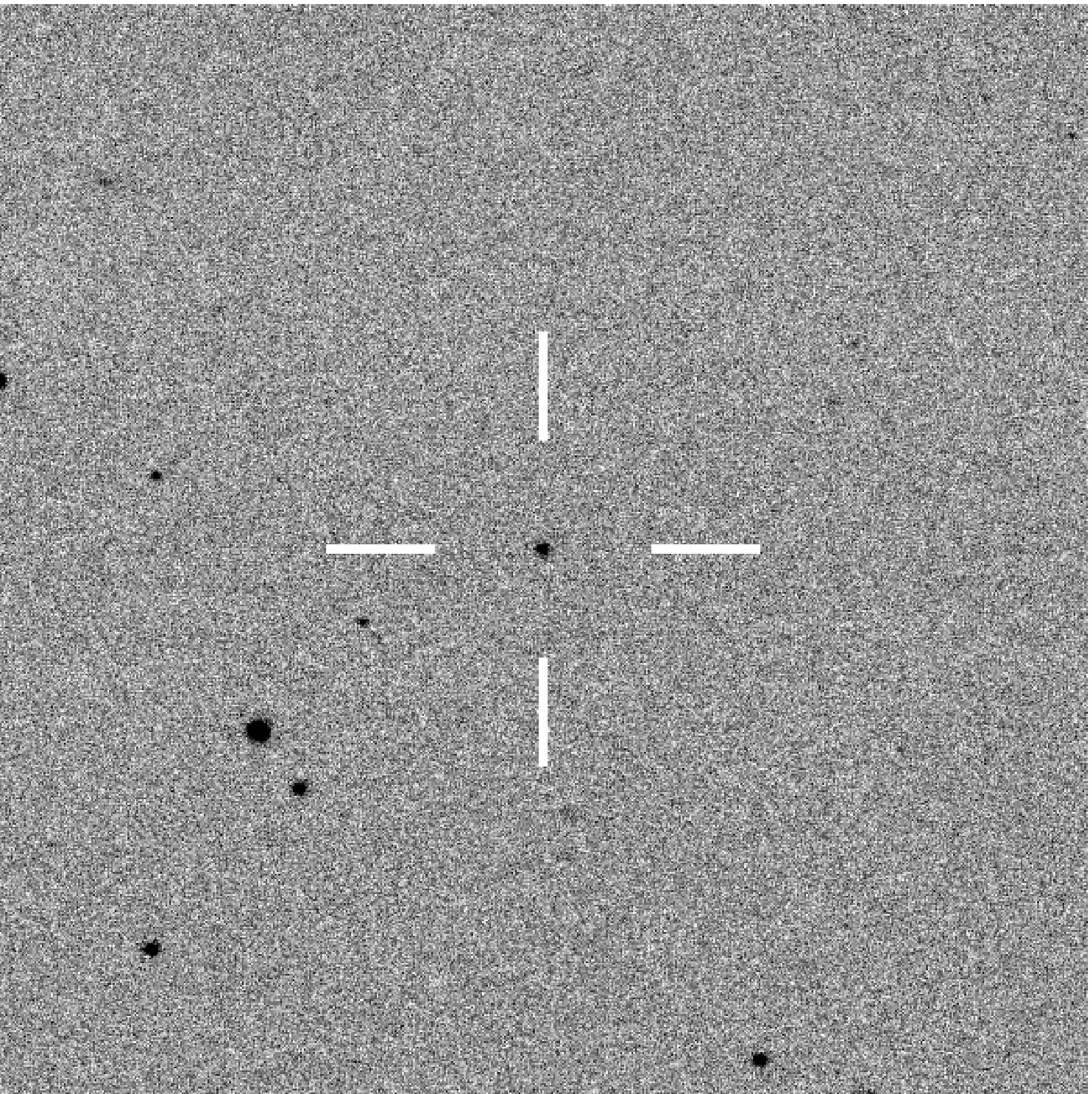}
   \includegraphics[height=26mm,angle=0]{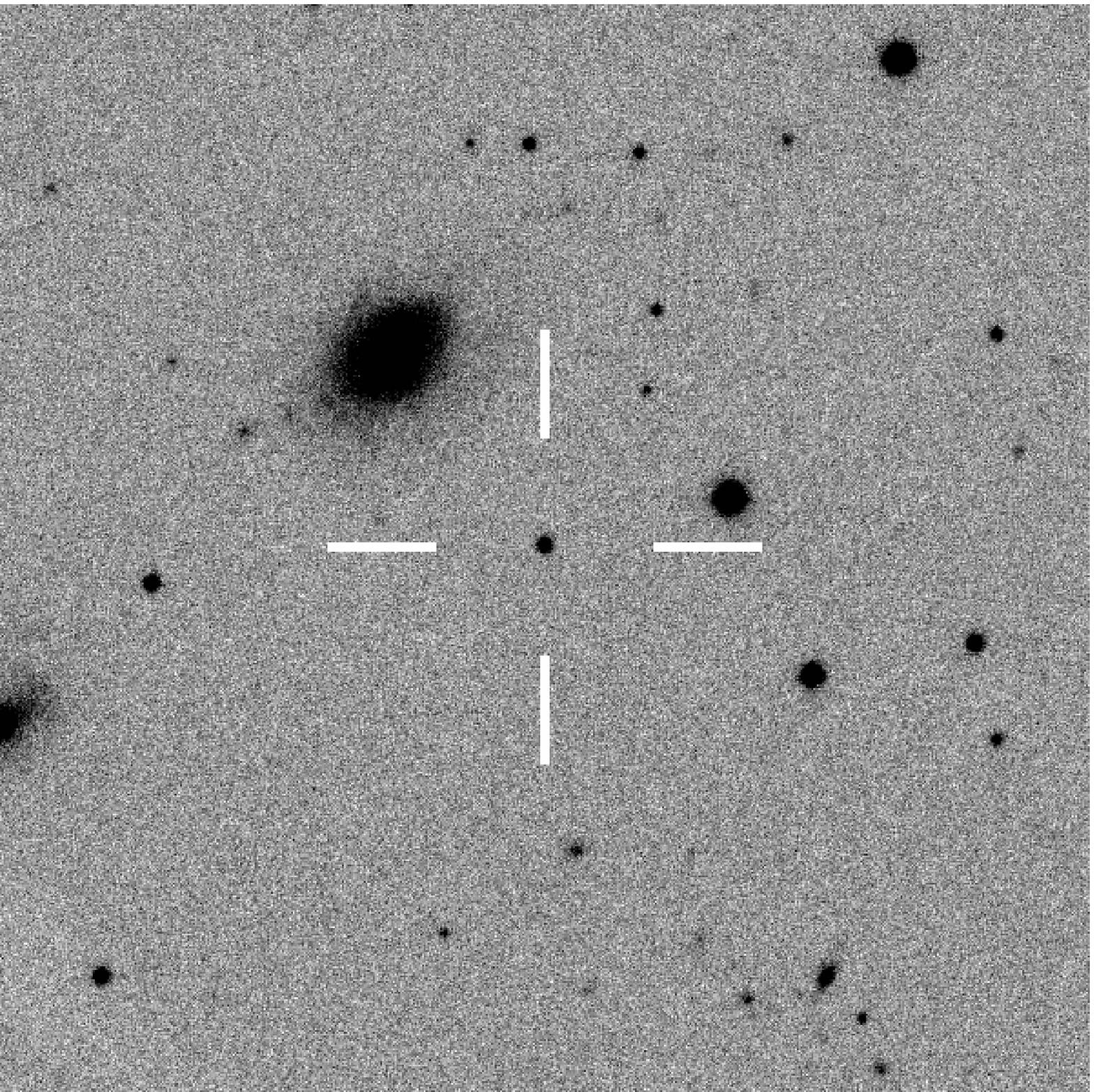}
   \includegraphics[height=26mm,angle=0]{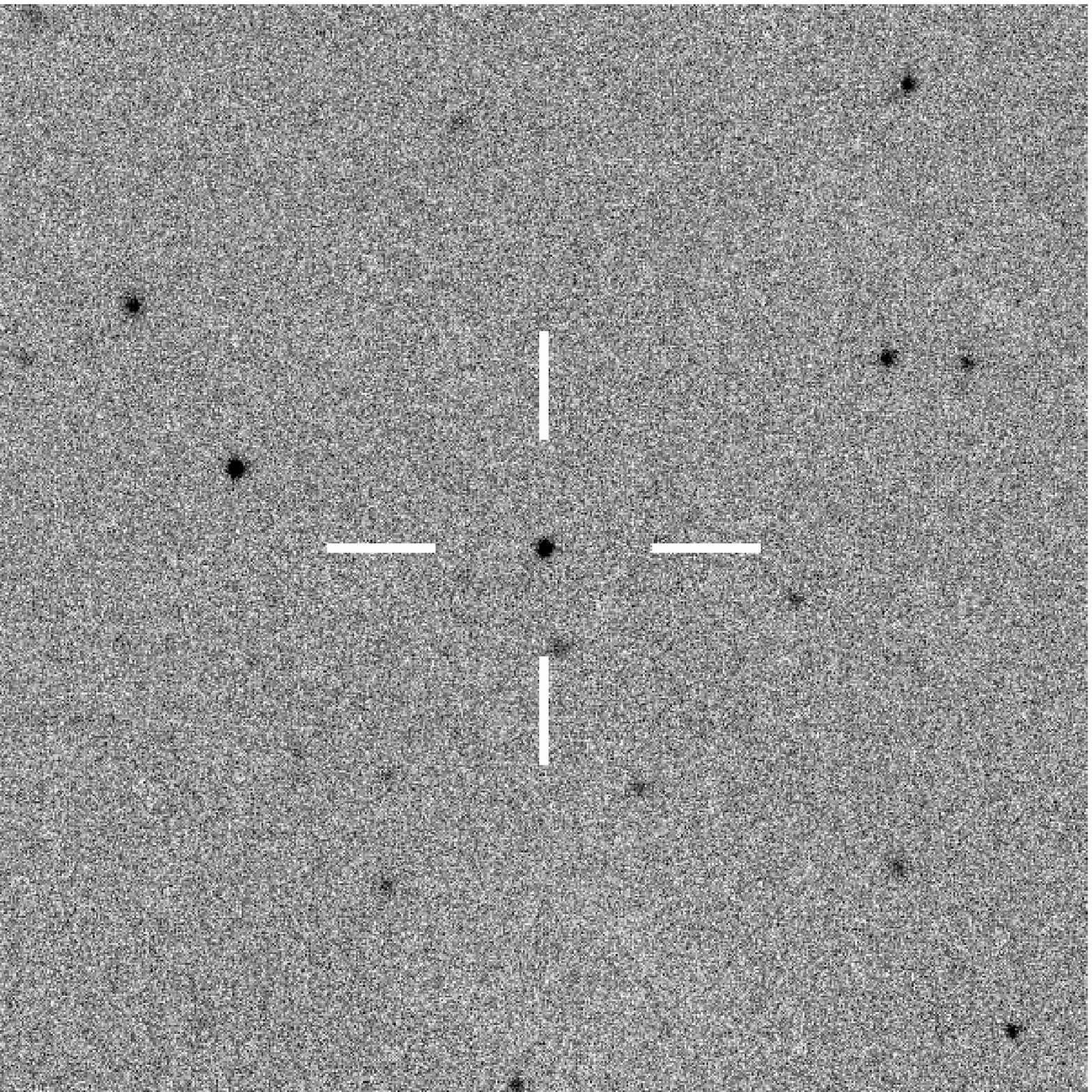}
   \includegraphics[height=26mm,angle=0]{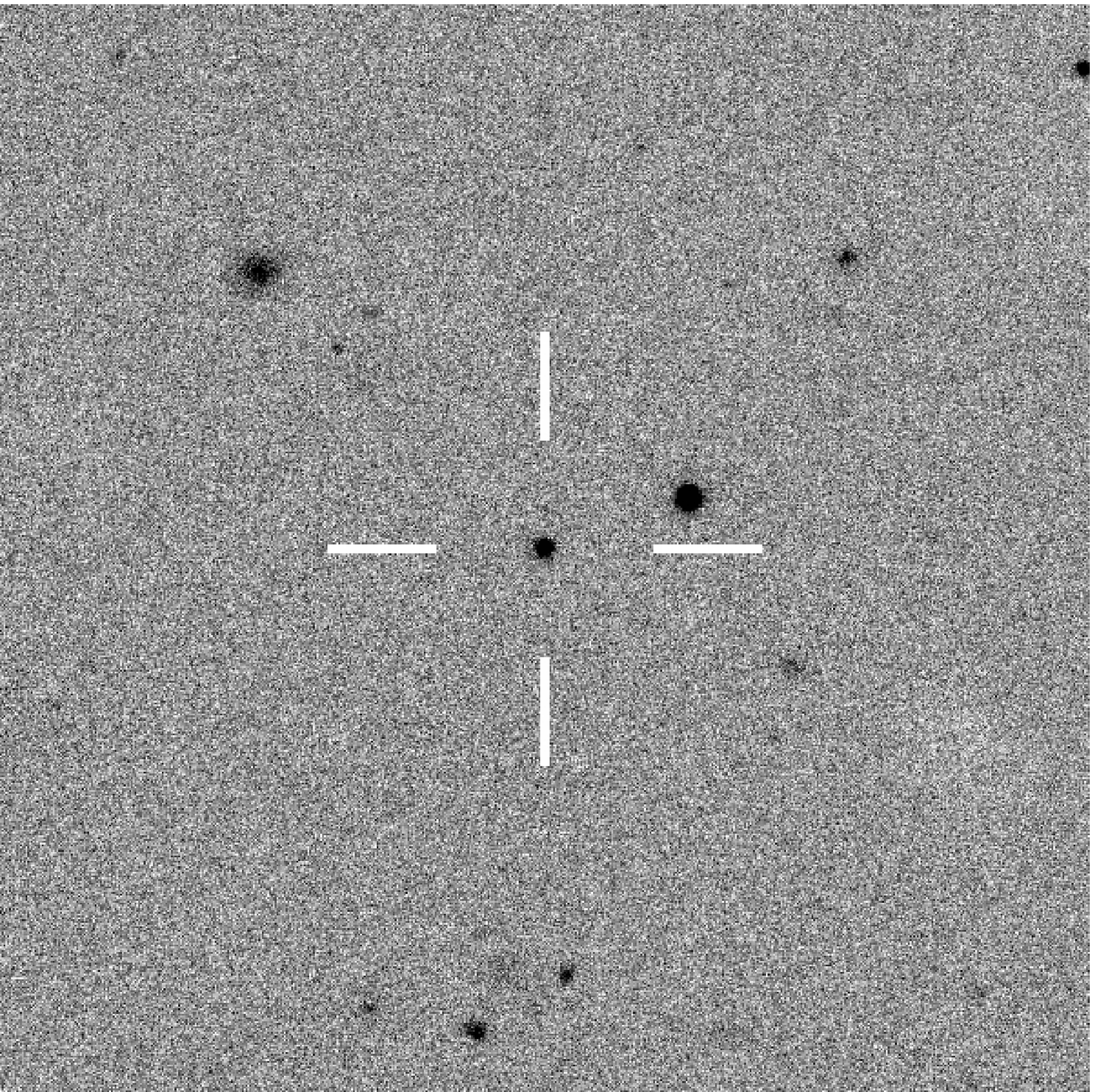}
   \includegraphics[height=26mm,angle=0]{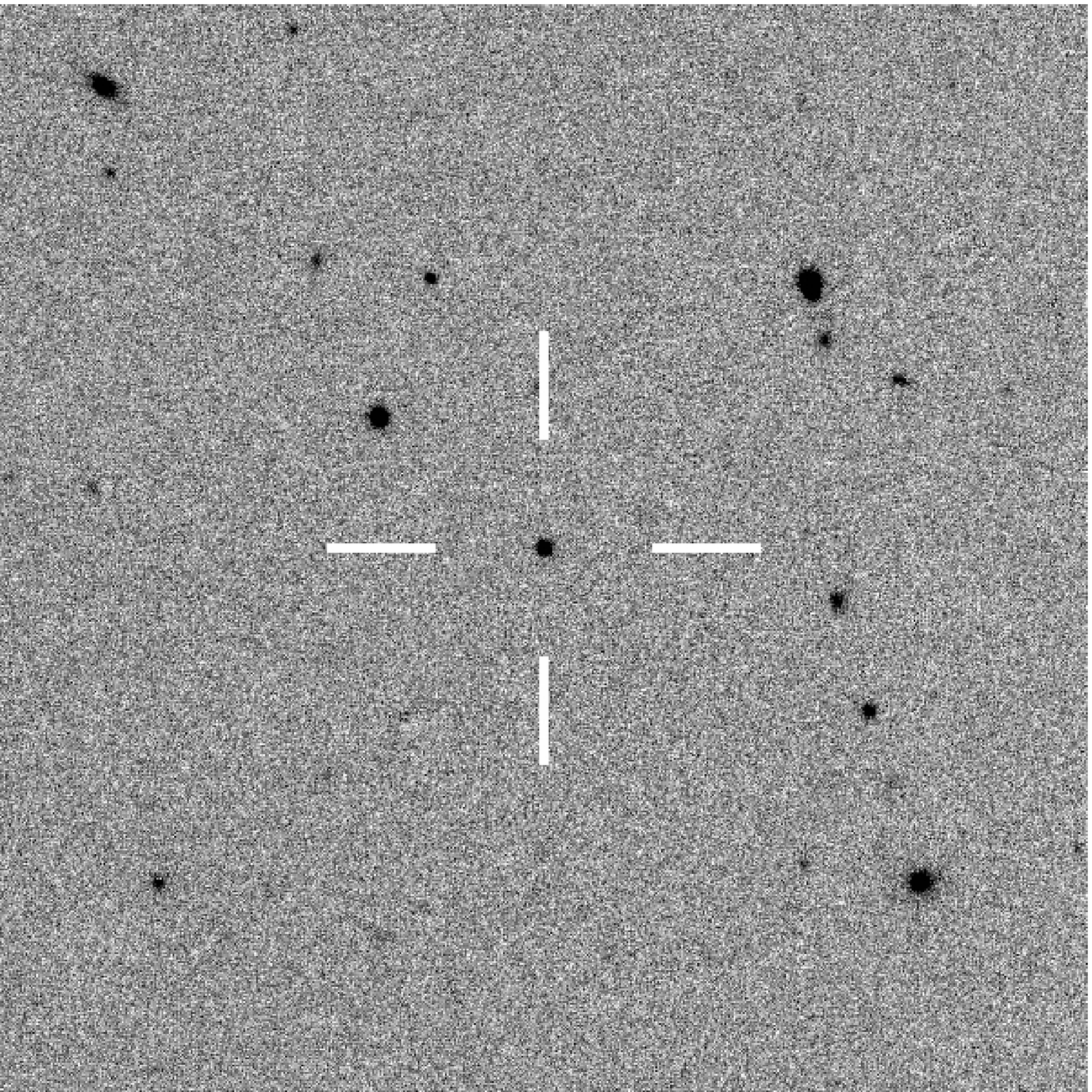}
   \includegraphics[height=26mm,angle=0]{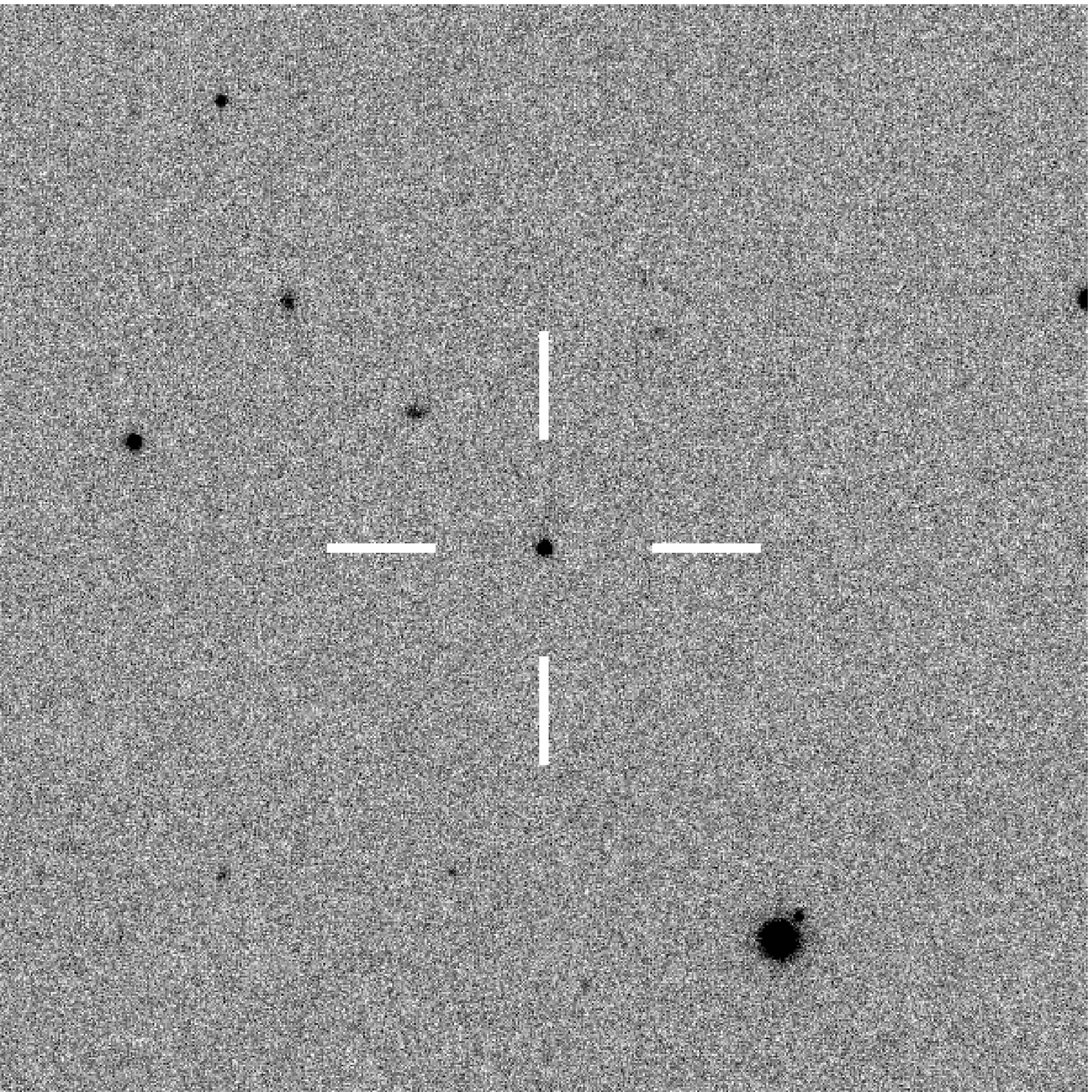}
   \includegraphics[height=26mm,angle=0]{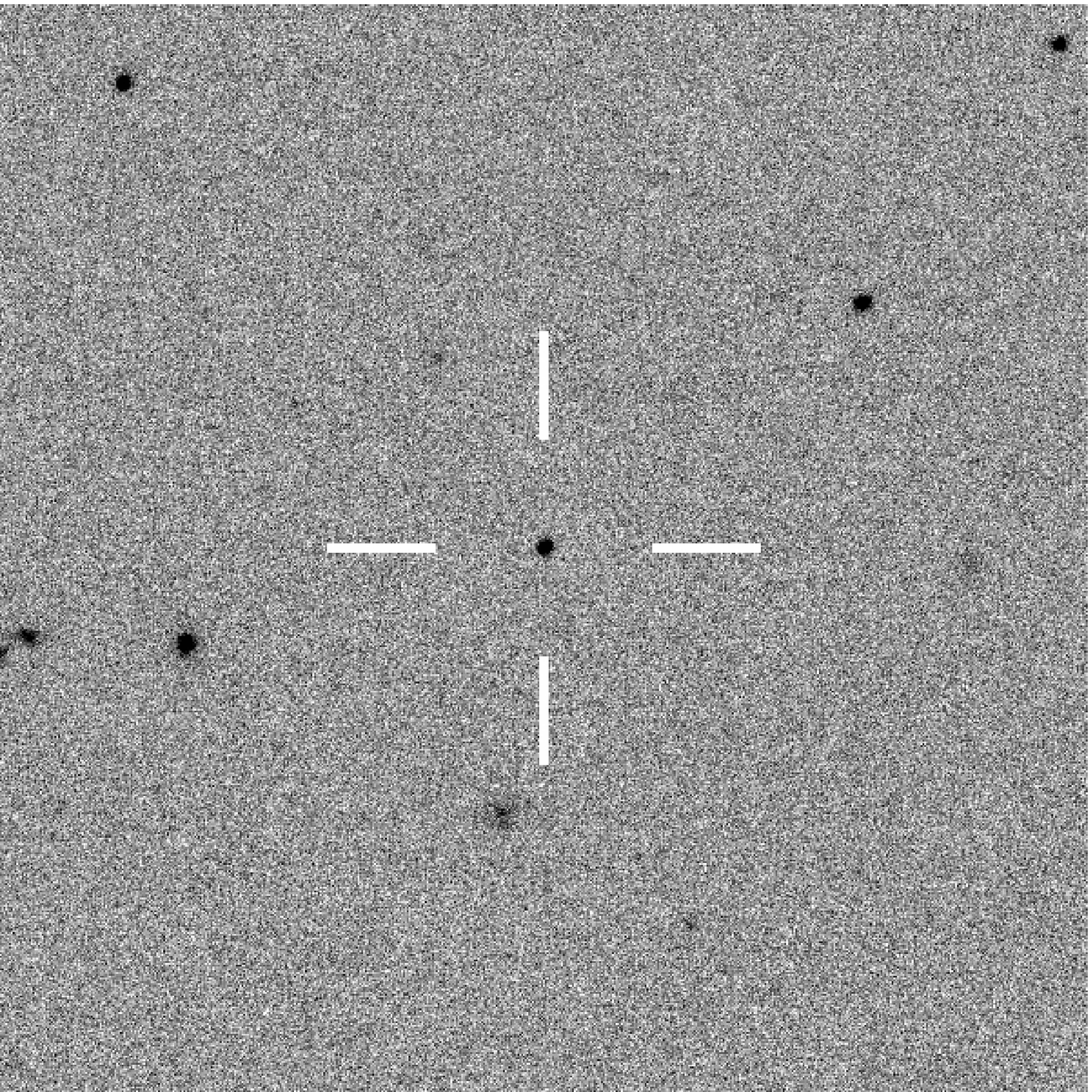}
   \includegraphics[height=26mm,angle=0]{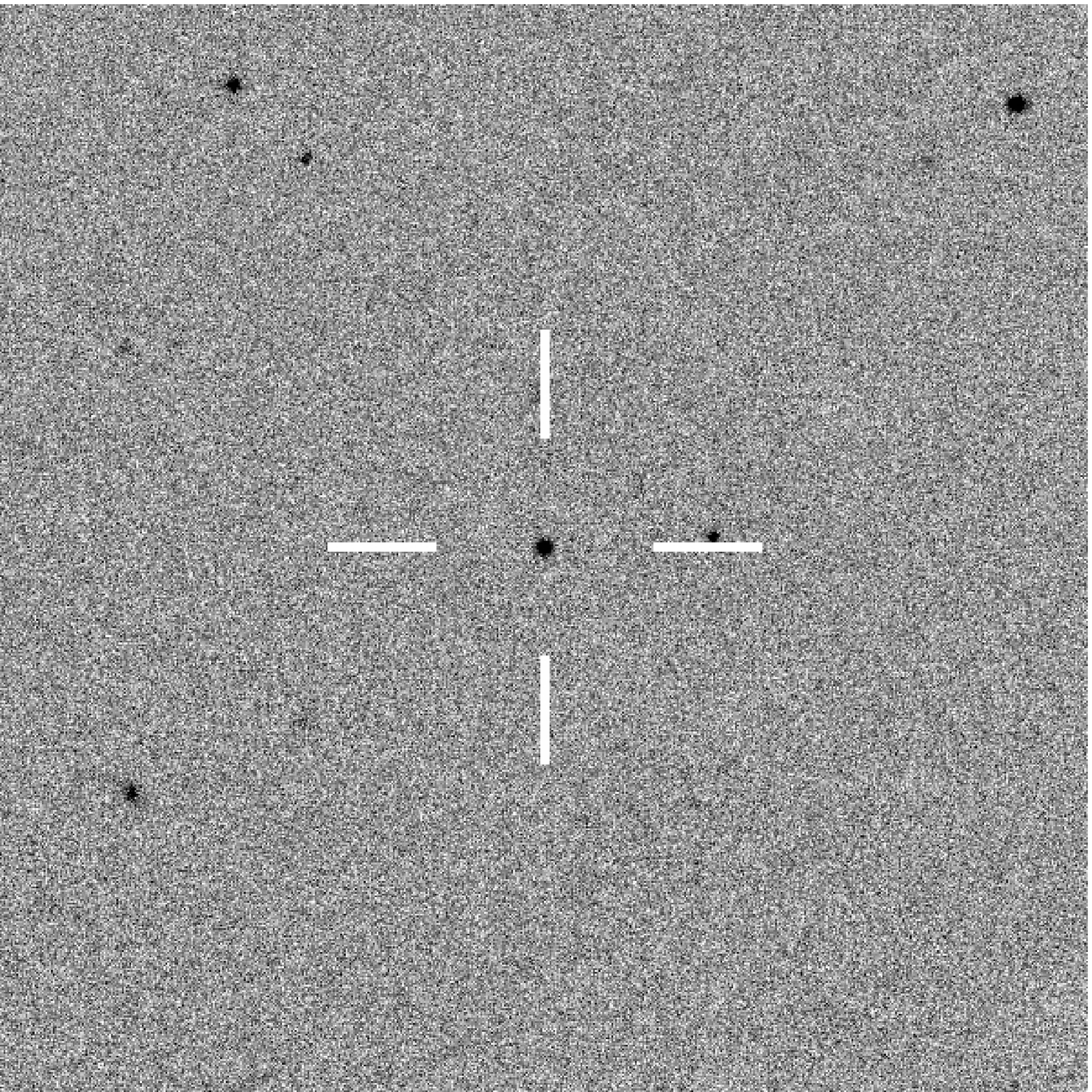}
   \includegraphics[height=26mm,angle=0]{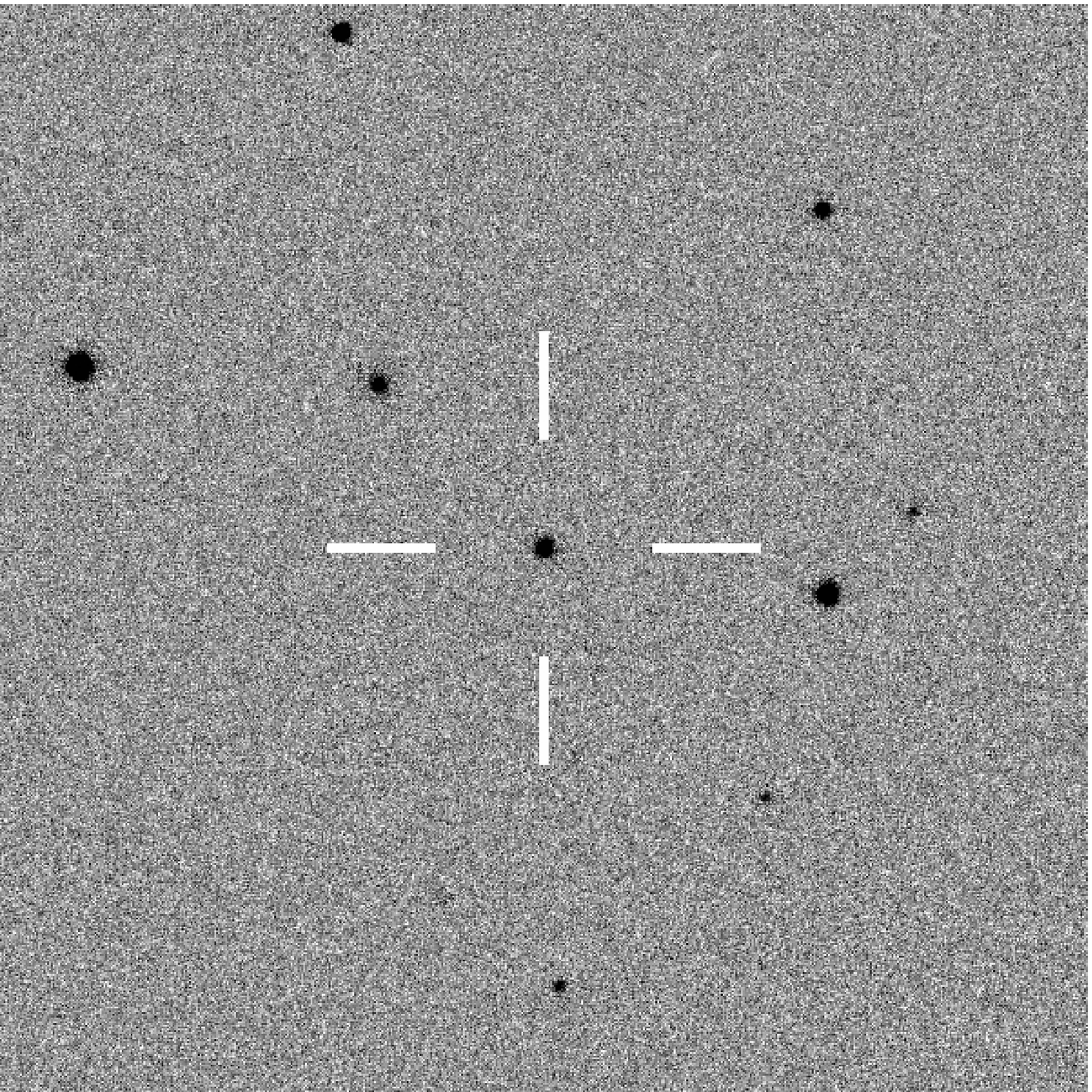}
      \caption{2$\times$2 arcmin$^2$ UKIDSS $J$-band images of 
              ULAS~J0329$+$04, ULAS~J0819$+$21, ULAS~J0945$+$07 (top row, from
              left to right), 
              ULAS~J1012$+$10, ULAS~J1417$+$13, ULAS~J1449$+$11 (middle row), 
              ULAS~J2320$+$14, ULAS~J2321$+$13, and ULAS~J2342$+$08 
              (bottom row). North is up, East to the left.
              }
         \label{fig_fcharts9}
   \end{figure}

\section{Astrometry \& photometry of T dwarf candidates}
\label{sect_pmphot}

Full names and photometry of 11 new T dwarf candidates 
are 
listed in Table~\ref{tab_photom}.
All candidates have $ugriz$ data where only the $z$ magnitudes are at or below
the SDSS detection limit, whereas the $ugri$ measurements 
are clearly above 
these limits. According to Abazajian et al.~(\cite{abazajian09}), 
a 95\% detection repeatability for point sources
is provided at $u$$=$22.0, $g$$=$22.2, $r$$=$22.2, $i$$=$21.3, $z$$=$20.5.

For the proper motion determination we took advantage of the sometimes 
slightly different epochs between the $YJHK$-band measurements (up to
one year epoch differences) and
of the additional epochs (separated by several years from the UKIDSS epochs)
provided by the SDSS (more details can be found in
Tables~\ref{tab_astm1} and \ref{tab_astm2}). For the proper motion 
solutions we used linear fitting over all available epochs combined
with a search for the correct counterparts at the expected positions
in data from additional epochs.

The errors of the proper motion components 
given in Tables~\ref{tab_pm1}, \ref{tab_pm2}, and \ref{tab_pm3}
are estimated from the scatter around the best-fit line in
each coordinate. 
and describe the quality of the fit.
Due to the small number of available epochs,
these formal errors may be unrealistically small 
(few data points may be well-aligned by chance)
compared to the
true errors. If we know the errors in the positions of the
objects measured 
at different epochs
in the SDSS and UKIDSS, we can estimate
which error in their proper motions would be expected for 
the given epoch differences
using error propagation.

The SDSS DR7 statistical errors per coordinate for bright stars 
are 45 mas, with systematic errors of less than 20~mas 
(Abazajian et al.~\cite{abazajian09}). However, for faint objects
at the SDSS survey limit, the astrometric accuracy is limited by 
photon statistics to about 100~mas (Pier et al.~\cite{pier03}).
The UKIDSS astrometric accuracy varies between 50~mas at low
galactic latitudes and 100~mas at high galactic latitudes
(Lawrence et al.~\cite{lawrence07}).
For our objects, mostly located at intermediate galactic latitudes
($|b|$=40$\degr$-50$\degr$), we assume
uncertainties of 100~mas and 70 mas 
respectively for single SDSS and UKIDSS positions, which 
include smaller colour-dependent systematic errors in the astrometry
expected to be of the order of a few tens of milliarcsec.
We combine all available SDSS data of a given object 
in a mean first epoch
and all UKIDSS data in a mean second epoch, where the accuracy
of the mean positions is getting higher with the square root 
of the number of available single epochs, and compute the
expected proper motion errors
from error propagation. 
These errors ($\varepsilon$) are given in
the last column of Tables~\ref{tab_pm1}, \ref{tab_pm2}, and \ref{tab_pm3}.
For the previously known T dwarfs (Table~\ref{tab_pm3}), available 2MASS
and DENIS positions were also assigned 100~mass individual errors and
used together with the SDSS positions for the mean first epoch, which
was then compared with the mean UKIDSS position to compute $\varepsilon$. 

The expected proper motion errors $\varepsilon$ are on average 
two times larger than the formal proper motion errors. Nevertheless,
all proper motions remain highly significant also compared to their
$\varepsilon$ values, except for two objects with proper motions 
only 3-5 times larger than the expected errors
(for further discussion, see Sect.~\ref{isoT}).
Note that the computed linear proper motions may be affected by 
parallactic motions for the nearest objects, since in most cases
the different epoch observations were not made during the same 
time of the year (see Tables~\ref{tab_astm1}, \ref{tab_astm2}).

\begin{table}
\caption{Multi-epoch positions $\alpha, \delta$ (J2000.0) of common proper motion
companions of Hipparcos stars}
\label{tab_astm1}
\centering
\begin{tabular}{lrllrl}
\hline\hline
\multicolumn{3}{c }{object} & epoch    & ID & source    \\
\multicolumn{1}{c}{$^s$} & \multicolumn{1}{c}{\arcmin} & \multicolumn{1}{c}{\arcsec} &  yr & &  \\
\hline
\multicolumn{5}{l }{ULAS~J1300$+$12 (Hip~63510C):} \\
41.9362 & 21 & 14.717 & 2003.223 & 3805 & SDSS DR7 \\
41.7711 & 21 & 14.754 & 2007.340 & 1465618 & UKIDSS $K$ \\
41.7709 & 21 & 14.747 & 2007.340 & 1467186 & UKIDSS $H$ \\
41.7286 & 21 & 14.695 & 2008.290 & 2321728 & UKIDSS $Y$ \\
41.7299 & 21 & 14.733 & 2008.290 & 2321748 & UKIDSS $J$ \\
\hline
\multicolumn{5}{l }{ULAS~J1504$+$05 ((Hip~73786B):}  \\
57.8614 & 38 & 03.278 & 2003.322 & 3910 & SDSS DR7 \\
57.6637 & 38 & 00.861 & 2008.151 & 1873794 & UKIDSS $H$ \\
57.6642 & 38 & 00.841 & 2008.151 & 1873842 & UKIDSS $K$ \\
57.6561 & 38 & 00.828 & 2008.293 & 2322073 & UKIDSS $Y$ \\
57.6561 & 38 & 00.849 & 2008.293 & 2322081 & UKIDSS $J$ \\
\hline
\end{tabular}

\smallskip

\scriptsize{
Note:
The ID is the run or multiframe number for SDSS or UKIDSS respectively.
}
\end{table}

\subsection{Common proper motion companions of Hipparcos stars}\label{cpmT}

After each successful proper motion determination we checked if the
newly found candidate is an already known T dwarf or if there are
other known stars with similar large proper motions within a search radius
of 2 degrees using SIMBAD. For two new candidates
we succeeded to find Hipparcos stars with very
similar proper motions (Table~\ref{tab_pm1}) based on which we
consider the two T dwarf candidates as wide companions of these
Hipparcos stars. ULAS~J1300$+$12 (Hip~63510C)
is separated by about 103~arcsec from Hip~63510A (Fig.~\ref{fig_fcharts2},
left panel) corresponding to a projected
physical separation of $\approx$1200~AU at 11.69$\pm$0.21~pc distance 
as derived from the newly determined Hipparcos parallax of 85.54$\pm$1.53~mas
(van Leeuwen~\cite{vanleeuwen07}). Hip~63510A (other names: Gl~494A;
LHS~2665A), an active, rapidly rotating M0.5 dwarf 
(see e.g. Rauscher \& Marcy~\cite{rauscher06};
Browning et al.~\cite{browning10}), 
has already a known close brown dwarf candidate companion 
with an astrometric orbit of 14.5 years and resolved by adaptive optics 
(Gl~494B; Heintz~\cite{heintz94};
Beuzit et al.~\cite{beuzit04}). For ULAS~J1504$+$05 ((Hip~73786B), the
separation from its primary, the K5 dwarf (Cenarro et al.~\cite{cenarro07})
Hip~73786A (other names: Gl~576A; LHS~3020A),
is about 68~arcsec (Fig.~\ref{fig_fcharts2}, right panel). This leads
to a projected physical separation of $\approx$1260~AU 
at 18.59$\pm$0.97~pc distance obtained
from the trigonometric parallax of 53.80$\pm$2.80~mas 
(van Leeuwen~\cite{vanleeuwen07}). 

\begin{table}
\caption{Common proper motions with Hipparcos stars in mas/yr}
\label{tab_pm1}
\centering
\begin{tabular}{l c c c}
\hline\hline
Object & $\mu_{\alpha}\cos{\delta}$ & $\mu_{\delta}$ & $\varepsilon$ \\
\hline
Hip~63510A & $-$616.3$\pm$1.5 &  $-$13.6$\pm$1.0 & \\
ULAS~J1300$+$12  & $-$596$\pm$7 &   $+$1$\pm$7 & 23\\
\hline
Hip~73786A & $-$608.8$\pm$2.7 & $-$502.7$\pm$3.6 & \\
ULAS~J1504$+$05  & $-$614$\pm$4 & $-$496$\pm$8 & 22\\
\hline
\end{tabular}

\smallskip

\scriptsize{
Notes:
Proper motions from linear fitting of the positions
given in Table~\ref{tab_astm1}. For Hipparcos primaries
data are taken from van Leeuwen~(\cite{vanleeuwen07}).
$\varepsilon$ are the expected errors (see text).
}
\end{table}

Goldman et al.~(\cite{goldman10}) have recently also reported on the discovery of
Hip~63510C (= Ross~458C). Their proper motion measurement is based on the UKIDSS
data and on additional new epoch observations with a total time baseline of two years
leading to larger uncertainties compared to our proper motion using a five years
baseline from SDSS and UKIDSS data.

\begin{table}
\caption{Multi-epoch positions $\alpha, \delta$ (J2000.0) of isolated objects}  
\label{tab_astm2}     
\centering            
\begin{tabular}{lrllrl}       
\hline\hline
\multicolumn{3}{c }{object} & epoch    & ID & source    \\
\multicolumn{1}{c}{$^s$} & \multicolumn{1}{c}{\arcmin} & \multicolumn{1}{c}{\arcsec} &  yr & &  \\
\hline
\multicolumn{5}{l }{ULAS~J0329$+$04:} \\
20.1535 & 30 & 24.390 & 2005.781 & 5714 & SDSS DR7 \\
20.1809 & 30 & 24.415 & 2006.816 & 6485 & SDSS DR7 \\
20.2211 & 30 & 24.558 & 2008.784 & 2336494 & UKIDSS $H$ \\
20.2227 & 30 & 24.538 & 2008.784 & 2336510 & UKIDSS $K$ \\
20.2170 & 30 & 24.549 & 2008.805 & 2338708 & UKIDSS $Y$ \\
20.2207 & 30 & 24.583 & 2008.805 & 2338724 & UKIDSS $J$ \\
\hline
\multicolumn{5}{l }{ULAS~J0819$+$21:} \\
18.5794 & 03 & 11.246 & 2004.209 & 4508 & SDSS DR7 \\
18.5888 & 03 & 10.809 & 2007.049 & 1184392 & UKIDSS $J$ \\
18.5848 & 03 & 10.480 & 2008.907 & 2340123 & UKIDSS $H$ \\
18.5792 & 03 & 10.497 & 2008.907 & 2340143 & UKIDSS $K$ \\
18.5849 & 03 & 10.441 & 2008.910 & 2341325 & UKIDSS $Y$ \\
18.5809 & 03 & 10.430 & 2008.910 & 2341345 & UKIDSS $J$ \\
\hline
\multicolumn{5}{l }{ULAS~J0945$+$07:} \\
16.1702 & 55 & 46.304 & 2002.194 & 3031 & SDSS DR7 \\
16.3889 & 55 & 45.586 & 2007.052 & 1186762 & UKIDSS $J$ \\
16.3974 & 55 & 45.632 & 2007.052 & 1188655 & UKIDSS $J$ \\
16.4008 & 55 & 45.652 & 2007.060 & 1343254 & UKIDSS $H$ \\
16.3905 & 55 & 45.650 & 2007.060 & 1343620 & UKIDSS $H$ \\
16.3977 & 55 & 45.636 & 2007.063 & 1330616 & UKIDSS $K$ \\
16.3937 & 55 & 45.589 & 2007.063 & 1332824 & UKIDSS $K$ \\
16.3947 & 55 & 45.584 & 2007.066 & 1335120 & UKIDSS $Y$ \\
16.3919 & 55 & 45.569 & 2007.066 & 1336848 & UKIDSS $Y$ \\
\hline
\multicolumn{5}{l }{ULAS~J1012$+$10:} \\
43.6322 & 21 & 04.108 & 2002.953 & 3538 & SDSS DR7 \\
43.5471 & 21 & 01.826 & 2007.005 & 1147818 & UKIDSS $K$ \\
43.5350 & 21 & 01.683 & 2007.249 & 1354063 & UKIDSS $H$ \\
43.5350 & 21 & 01.698 & 2007.304 & 1415926 & UKIDSS $J$ \\
43.5356 & 21 & 01.698 & 2007.304 & 1417555 & UKIDSS $Y$ \\
\hline
\multicolumn{5}{l }{ULAS~J1417$+$13:} \\
56.2478 & 30 & 45.565 & 2003.472 & 3996 & SDSS DR7 \\
56.2275 & 30 & 45.875 & 2007.367 & 1450483 & UKIDSS $Y$ \\
56.2263 & 30 & 45.868 & 2007.367 & 1450579 & UKIDSS $J$ \\
56.2226 & 30 & 45.943 & 2008.222 & 2318417 & UKIDSS $K$ \\
56.2236 & 30 & 45.915 & 2008.222 & 2318429 & UKIDSS $H$ \\
\hline
\multicolumn{5}{l }{ULAS~J1449$+$11:} \\
01.9706 & 47 & 12.307 & 2003.409 & 3971 & SDSS DR7 \\
01.9050 & 47 & 11.390 & 2007.260 & 1367513 & UKIDSS $J$ \\
01.9064 & 47 & 11.415 & 2007.260 & 1368654 & UKIDSS $Y$ \\
01.9138 & 47 & 11.267 & 2007.293 & 1402270 & UKIDSS $K$ \\
01.9024 & 47 & 11.389 & 2007.293 & 1402702 & UKIDSS $H$ \\
\hline
\multicolumn{5}{l }{ULAS~J2320$+$14:} \\
35.1020 & 48 & 28.996 & 2000.735 & 1739 & SDSS P \\
35.1064 & 48 & 28.978 & 2000.910 & 1904 & SDSS P \\
35.2839 & 48 & 29.835 & 2007.729 & 1716958 & UKIDSS $K$ \\
35.2900 & 48 & 29.815 & 2007.729 & 1716859 & UKIDSS $H$ \\
35.2914 & 48 & 29.815 & 2007.729 & 1717061 & UKIDSS $Y$ \\
35.2890 & 48 & 29.836 & 2007.729 & 1717217 & UKIDSS $J$ \\
\hline
\multicolumn{5}{l }{ULAS~J2321$+$13:} \\
23.7591 & 54 & 59.058 & 2000.735 & 1739 & SDSS P \\
23.7724 & 54 & 58.799 & 2000.910 & 1904 & SDSS P \\
23.7910 & 54 & 58.261 & 2001.639 & 2507 & SDSS P \\
23.7990 & 54 & 54.895 & 2007.666 & 1689556 & UKIDSS $J$ \\
23.7990 & 54 & 54.943 & 2007.666 & 1690721 & UKIDSS $K$ \\
23.7990 & 54 & 54.918 & 2007.666 & 1690865 & UKIDSS $Y$ \\
23.7969 & 54 & 54.935 & 2007.666 & 1691223 & UKIDSS $H$ \\
\hline
\multicolumn{5}{l }{ULAS~J2342$+$08:} \\
28.9358 & 56 & 20.123 & 2006.715 & 6354 & SDSS DR7 \\
28.9455 & 56 & 20.133 & 2007.901 & 1830486 & UKIDSS $H$ \\
28.9423 & 56 & 20.140 & 2007.901 & 1830585 & UKIDSS $K$ \\
28.9659 & 56 & 20.111 & 2008.732 & 2335155 & UKIDSS $Y$ \\
28.9655 & 56 & 20.105 & 2008.732 & 2335171 & UKIDSS $J$ \\
\hline
\end{tabular}

\smallskip

\scriptsize{
Note:
The ID is the run or multiframe number for SDSS or UKIDSS respectively. 
SDSS P stands for the SDSS data base in Princeton.
}
\end{table}

\begin{table}
\caption{Proper motions of isolated objects in mas/yr}
\label{tab_pm2}
\centering
\begin{tabular}{l r r r}
\hline\hline
Object & $\mu_{\alpha}\cos{\delta}$ & $\mu_{\delta}$ & $\varepsilon$ \\
\hline
ULAS~J0329$+$04 & $+$323$\pm$15 &   $+$60$\pm$07 & 32 \\ 
ULAS~J0819$+$21 &   $+$5$\pm$14 &  $-$168$\pm$07 & 24 \\
ULAS~J0945$+$07 & $+$685$\pm$13 &  $-$142$\pm$07 & 21 \\
ULAS~J1012$+$10 & $-$327$\pm$11 &  $-$558$\pm$06 & 25 \\
ULAS~J1417$+$13 &  $-$76$\pm$03 &   $+$77$\pm$03 & 24 \\
ULAS~J1449$+$11 & $-$242$\pm$21 &  $-$244$\pm$18 & 27 \\
ULAS~J2320$+$14 & $+$387$\pm$05 &  $+$121$\pm$02 & 11 \\
ULAS~J2321$+$13 &  $+$56$\pm$15 &  $-$577$\pm$10 & 10 \\
ULAS~J2342$+$08 & $+$229$\pm$55 &    $-$9$\pm$09 & 48 \\
\hline
\end{tabular}

\smallskip

\scriptsize{
Notes:
Proper motions come from linear fitting of the positions
given in Table~\ref{tab_astm2}.
$\varepsilon$ are the expected errors (see text).
}
\end{table}

\begin{table}
\caption{Proper motions of known T dwarfs in mas/yr}
\label{tab_pm3}
\centering
\begin{tabular}{@{\extracolsep{-5pt}}llrrr}
\hline\hline
Object & SpT (Ref) & $\mu_{\alpha}\cos{\delta}$ & $\mu_{\delta}$ & $\varepsilon$ \\
\hline
SDSS~J0325$+$04     & T5.5 (1) &  $-$185$\pm$06 &   $+$15$\pm$07 & 27 \\
2MASSI~J0755$+$22   & T5.0 (2) &   $-$13$\pm$03 &  $-$241$\pm$05 & 11 \\
SDSS~J0830$+$01     & T4.5 (2) &  $+$214$\pm$06 &  $-$329$\pm$09 & 08 \\
2MASS~J1231$+$08    & T5.5 (2) & $-$1185$\pm$09 & $-$1050$\pm$12 & 14 \\
SDSS~J1504$+$10     & T7.0 (1) &  $+$338$\pm$13 &  $-$350$\pm$11 & 21 \\
2MASSI~J2339$+$13   & T5.0 (2) &  $+$380$\pm$05 &  $-$971$\pm$05 & 12 \\
\hline
\end{tabular}

\smallskip

\scriptsize{
Notes:
Proper motions as obtained from linear fitting of 
UKIDSS, SDSS, and
(if available) 2MASS and DENIS positions. 
$\varepsilon$ are the expected errors (see text).
For full names
and references, see Gelino et al.~(\cite{gelino09}).
Spectral type (SpT) references (Ref):
(1) - Chiu et al.~(\cite{chiu06}),
(2) - Burgasser et al.~(\cite{burgasser06b}).
}
\end{table}

\subsection{Isolated T dwarf candidates}\label{isoT}

The individual astrometric measurements of the remaining nine candidates,
which appeared to have no common proper motion objects, are listed in
Table~\ref{tab_astm2}. Their proper motions are given in Table~\ref{tab_pm2}.
In most cases, there are more than just two effective epochs available,
leading to a reliable linear proper motion fit.
This is also true for the object with the smallest proper motion,
ULAS~J1417$+$13, which shows however the largest discrepancy between the
very small formal proper motion errors and the expected proper motion
error $\varepsilon$. But even compared to the latter, both proper motion
components are three times larger, so significant.
The reason for the relatively
uncertain proper motion 
(4-5 times larger than its error)
in case of ULAS~J2342$+$08 is the late SDSS epoch
and resulting short total time baseline (only 2 years).

\section{Proper motions of previously known T dwarfs}
\label{sect_knownT}

In Table~\ref{tab_pm3}, the proper motions of six known T dwarfs
from the compilation of Gelino et al.~(\cite{gelino09})
detected in our combined survey of UKIDSS and SDSS are listed.
In part, they were determined using additional epoch data from
2MASS and DENIS.

For four of the known T dwarfs, these are the first proper
motion measurements. For 2MASS~J1231$+$08
and 2MASSI~J2339$+$13, we have
obtained similar but
five to ten times
more accurate proper motions than
determined earlier by Burgasser et al.~(\cite{burgasser04})
and Burgasser et al.~(\cite{burgasser03}) respectively,
even if we prefer the expected proper motion errors
$\varepsilon$ to our formal errors.

\section{Conclusions and discussion}~\label{res}

Whereas for three of our candidates we measure moderate $z$$-$$J$=$+$2.6...$+$2.8
meeting our selection criteria,
eight objects have much larger values of $z$$-$$J$=$+$3.3...$+$3.8 (ULAS~J0819$+$21
has $z$$-$$J$$\approx$$+$3.8, ULAS~J1417$+$13 and
ULAS~J2342$+$08 have $z$$-$$J$$\approx$$+$3.6)
approaching the typical values of T8-T9 dwarfs
(Warren et al.~\cite{warren07}; Burningham et al.~\cite{burningham08,burningham09};
Delorme et al~\cite{delorme08a}).

%
\begin{table*}
\caption{Estimated spectral types of T dwarf candidates based on colours and absolute magnitudes}
\label{tab_expspt}
\centering
\begin{tabular}{l l l l l l}
\hline\hline
Object & SpT $(J-H)$ & SpT $(J-K)$& SpT $(M_J)$ & SpT $(M_K)$ & SpT adopted \\
\hline
ULAS~J0329$+$04 &  T5.0-T9.0 ($-$0.38) & T5.5-T9.0 ($-$0.67) &                            &                            & T7.0$\pm$2.0 \\
ULAS~J0819$+$21 &  T5.0-T6.5 ($-$0.33) & T4.0-T6.5 ($-$0.23) &                            &                            & T5.5$\pm$1.0 \\
ULAS~J0945$+$07 &  T4.5-T6.0 ($-$0.22) & T4.0-T7.5 ($-$0.30) &                            &                            & T5.5$\pm$1.5 \\
ULAS~J1012$+$10 &  T5.0-T9.0 ($-$0.37) & T5.5-T9.0 ($-$0.57) &                            &                            & T7.0$\pm$2.0 \\
Hip~63510C      &  T5.0-T6.5 ($-$0.33) & T4.0-T6.5 ($-$0.22) & T7.5-T8.0 (16.35) & T6.5-T8.0 (16.56) & T7.0$\pm$1.0 \\
ULAS~J1417$+$13 &  T5.0-T5.5 ($-$0.23) & T4.0-T6.5 ($-$0.23) &                            &                            & T5.5$\pm$1.0 \\
ULAS~J1449$+$11 &  T5.0-T9.0 ($-$0.37) & T5.5-T9.0 ($-$0.75) &                            &                            & T7.0$\pm$2.0 \\
Hip~73786B      &  T6.5-T9.0 ($-$0.46) & T6.0-T9.0 ($-$0.82) & T6.0-T7.5 (15.25) & T6.5-T7.5 (16.07) & T6.5$\pm$1.0 \\ 
ULAS~J2320$+$14 &  T5.0-T9.0 ($-$0.34) & T5.0-T9.0 ($-$0.49) &                            &                            & T7.0$\pm$2.0 \\
ULAS~J2321$+$13 &  T5.5-T9.0 ($-$0.40) & T5.5-T9.0 ($-$0.67) &                            &                            & T7.5$\pm$1.5 \\
ULAS~J2342$+$08 &  T5.0-T9.0 ($-$0.36) & T5.5-T9.0 ($-$0.61) &                            &                            & T7.0$\pm$2.0 \\
\hline
\end{tabular}
\end{table*}

   \begin{figure}
   \centering
   \includegraphics[height=81mm,angle=270]{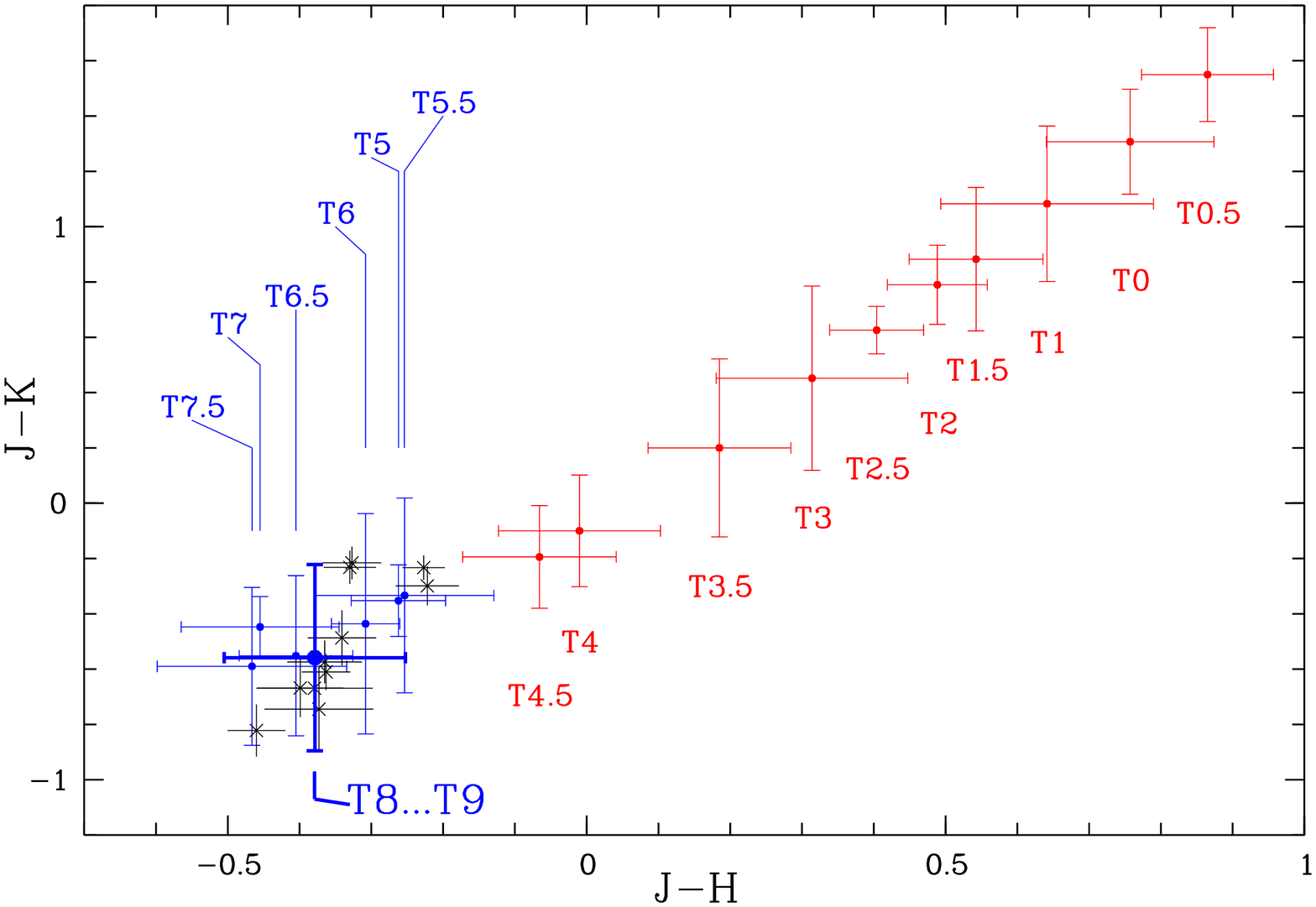}
   \includegraphics[height=81mm,angle=270]{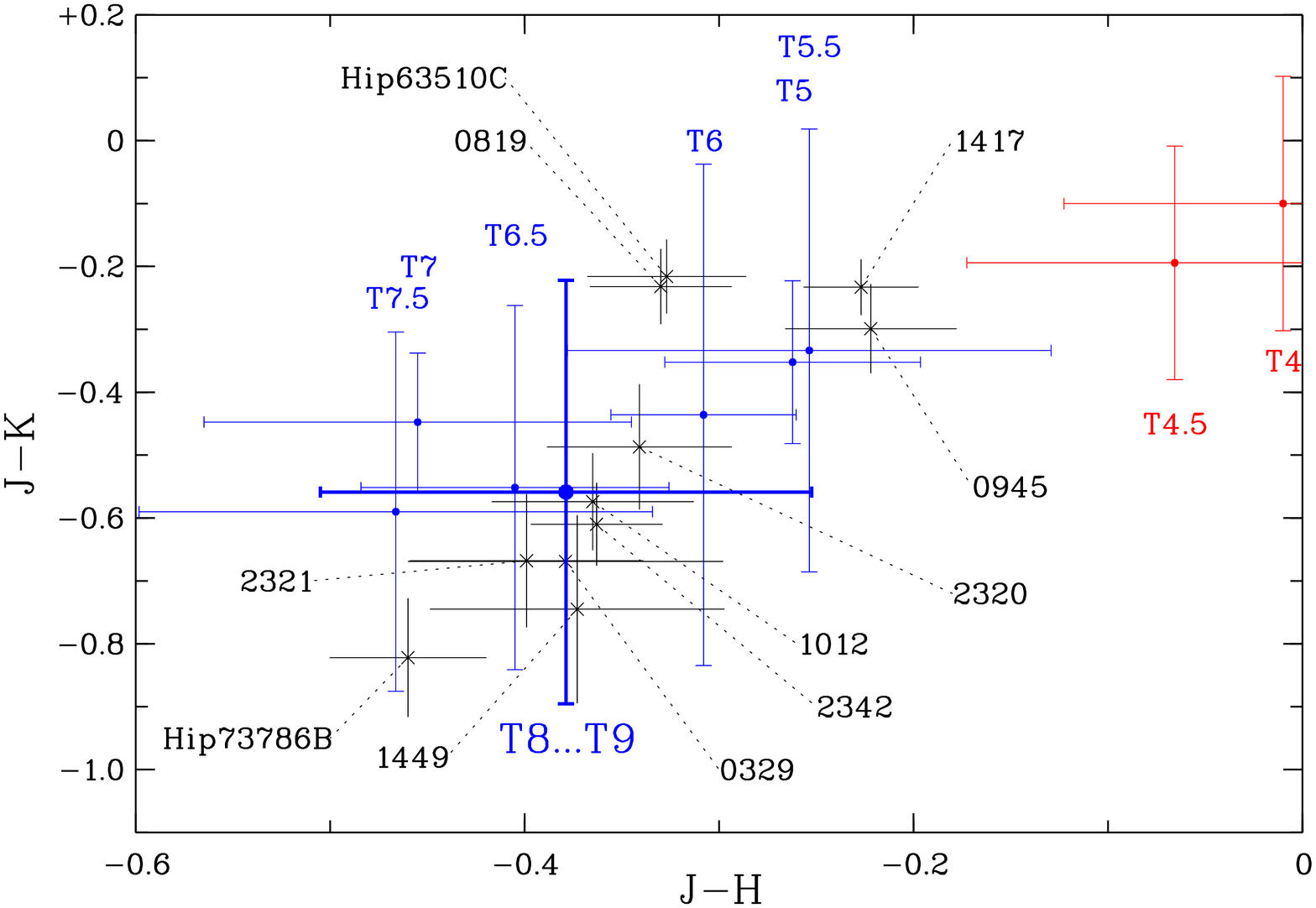}
      \caption{Upper panel: Mean $J$$-$$K$ vs. $J$$-$$H$ (filled symbols)
               in MKO system for different sub-classes of
               T dwarfs (labelled; their error bars represent standard deviations)
               from Leggett et al.~(\cite{leggett10}) with the colours
               of our 11 candidates overplotted as crosses with error bars.
               Lower panel: Zoom to late T dwarf region,
               where the targets are also labelled.
              }
         \label{fig_2col}
   \end{figure}

The mean near-infrared colours $J$$-$$K$ and $J$$-$$H$ of known T dwarfs
follow a clear trend to bluer colour with later sub-type (Fig.~\ref{fig_2col}),
except for T8-T9 dwarfs which again show colours similar to T6-T7.5 dwarfs.
All our new T dwarf candidates exhibit sufficiently blue colours classifying
them as $\ge$T5 dwarfs. Based on these colours we assign a preliminary
classification as 
$\approx$T5.5 (with uncertainties up to $\pm$1.5 subtypes)
for ULAS~J0819$+$21, ULAS~J0945$+$07,
ULAS~J1300$+$12 (Hip~63510C), and ULAS~J1417$+$13 (both $J$$-$$K$
and $J$$-$$H$ are between -0.2 and -0.35). For the remaining objects
with even larger negative colour indices,
including ULAS~J1504$+$05 (= Hip~73786B) with the most extreme
values $J$$-$$K$$\approx$$-$0.8 and $J$$-$$H$$\approx$$-$0.5,
the colours indicate T5-T9 
spectral types 
(Table~\ref{tab_expspt}). The larger uncertainty comes here mainly from
the already mentioned colour reversal of T8-T9 dwarfs.
 
Absolute magnitudes of the two Hipparcos star companions can be derived
using the accurate distances of their primaries. Taking into account the errors
in the distances and apparent magnitudes, we get $M_J$ of 16.35$\pm$0.04
and 15.25$\pm$0.12, and $M_K$ of 16.56$\pm$0.07 and 16.07$\pm$0.15 respectively 
for Hip~63510C and Hip~73786B. These absolute magnitudes place Hip~63510C
among the 
T6-T8 
dwarfs with trigonometric parallaxes as given in
Leggett et al.~(\cite{leggett10}), whereas Hip~73786B is as faint as 
T5.5-T7.5 
dwarfs listed in that paper. The fact that the colours hint at
a later spectral type for Hip~73786B compared to Hip~63510C, whereas the
classification by absolute magnitudes preferred by us
lead to a different conclusion, underlines the
need for spectroscopic observations of our new targets and direct comparison 
with template spectra of the coolest known brown dwarfs. 
Goldman et al.~(\cite{goldman10}) discuss Hip~63510C in more detail
and come to the conclusion that it is a T8-T9 dwarf with an absolute magnitude
and $J-K$ colour pointing to a young age or possible binarity.

\begin{acknowledgements}
Data from the UKIDSS 6th data release, SDSS DR7, 2MASS, and DENIS 
served as the basis for this work. We have also used the SIMBAD and 
VizieR services at the CDS in Strasbourg and the ARICNS data base on nearby
stars in Heidelberg. We thank Doug Finkbeiner for his
help accessing the SDSS data at Princeton University and Axel Schwope
for comments on a first version of the manuscript. 
We also thank the referee, Dr. P. Delorme, for helpful 
comments which led to an improved and extended paper.
\end{acknowledgements}

\end{document}